\documentstyle[here,epsfig]{mn2e}
\def\simlt{\mathrel{\rlap{\lower 3pt\hbox{$\sim$}}\raise 2.0pt\hbox{$<$}}}
\def\simgt{\mathrel{\rlap{\lower 3pt\hbox{$\sim$}} \raise 2.0pt\hbox{$>$}}}

\def\gtsima{$\; \buildrel > \over \sim \;$}
\def\ltsima{$\; \buildrel < \over \sim \;$}
\def\gtrsim{\lower.5ex\hbox{\gtsima}}
\def\lesssim{\lower.5ex\hbox{\ltsima}}

\newcommand{\q}{\begin{equation}}
\newcommand{\qa}{\begin{eqnarray}}
\newcommand{\qs}{\begin{eqnarray*}}
\newcommand{\nq}{\end{equation}}
\newcommand{\nqa}{\end{eqnarray}}
\newcommand{\nqs}{\end{eqnarray*}}

\begin{document}

\def\kms{{\rm\,km\,s^{-1}}}
\def\persec{{\rm s^{-1}}}
\def\gcm3{{\rm g\,\, cm^{-3}}}
\def\msun{M_\odot}
\def\pc3{{\rm pc}^{-3}}
\def\vecJstar{{\bf {J}}_{\ast{}}}
\def\vecJBH{{\bf {J}}_{\rm {BH}} }
\def\Jstar{J_{\ast{}}}
\def\JBH{J_{\rm {BH}}}

\title[The fingerprint of binary intermediate mass black holes]{The fingerprint of binary intermediate mass black holes in  globular clusters: supra-thermal stars and angular momentum alignment}

\author[M. Mapelli, M. Colpi, A. Possenti \& S. Sigurdsson]{
M. Mapelli$^{1}$,
M. Colpi$^{2}$, A. Possenti$^{3}$
\&  S. Sigurdsson$^{4}$\\
$^{1}$S.I.S.S.A., Via Beirut 2 - 4, I-34014 Trieste, Italy; {\tt mapelli@sissa.it}\\
$^{2}$Dipartimento di Fisica G. Occhialini, Universit\`a di
Milano Bicocca, Piazza della Scienza 3. I-20126 Milano, Italy; \\{\tt monica.colpi@mib.infn.it}\\
$^{3}$INAF, Osservatorio Astronomico di Cagliari, Poggio dei Pini, Strada 54, 09012 Capoterra, Italy; {\tt possenti@ca.astro.it}\\
$^{4}$Department of Astronomy and Astrophysics, The Pennsylvania State
University, 525 Davey Lab, University Park, PA~16802; \\{\tt steinn@astro.psu.edu}
}

\maketitle \vspace {7cm }

\begin{abstract}
We explore the signatures that a binary intermediate mass black hole (IMBH)
imprints on the velocity
and on the angular momentum of globular cluster stars. Simulating
3-body encounters between a star and a binary IMBH, we find that the
binary IMBH generates a family of few hundreds of stars
($\sim{}$100-300) which remain bound to the globular cluster (GC) and have velocity
significantly higher than the dispersion velocity. For this reason we
term them ''supra-thermal'' stars.  We also notice that, after the
interaction, a considerable fraction (55-70\%) of stars tend to align
their orbital angular momentum with the angular momentum of the binary
IMBH, introducing an anisotropy in the angular momentum distribution
of cluster stars. 
We simulate the
dynamical evolution of these supra-thermal stars before
thermalization, and find that these stars tend to cluster at a
distance of few core radii from the GC center. 
We conclude that the detectability of such signatures appears problematic 
with present telescopes.


\end{abstract}


\section{Introduction}

A number of different observations suggest that large black holes
(BHs) may exist in nature, with masses between $20\msun - 10^4
M_{\odot}.$ Heavier than the stellar-mass BHs born in core-collapse
supernovae ($3\msun-20\msun$; Orosz 2002), these intermediate mass
black holes (IMBHs) are expected to form from the direct collapse of
very massive stars, or in dense stellar systems through complex
dynamical processes. Filling the gap between stellar-mass black holes
(BHs) and super-massive black holes (SMBHs), they are of crucial
importance in establishing the potential physical link between these
two classes.

IMBHs may plausibly have formed in the early universe as remnants of
the first generation of metal free stars (Abel, Bryan \& Norman 2002;
Heger et al. 2003; Schneider et al. 2002).  If this is true, IMBHs can
participate the cosmic assembly of galaxies and be incorporated in
larger and larger units, becoming seeds for the formation of SMBHs
(Madau \& Rees 2001; Volonteri, Haardt \& Madau 2003).  IMBH may still
be forming in young dense star clusters vulnerable to unstable mass
segregation, via collisions of massive stars (Portegies Zwart 2004;
Portegies Zwart et al. 2004; Portegies Zwart \& McMillan 2002; G\"urkan et al. 2004; Freitag et al. 2005a, 2005b; see van
der Marel 2004 for a review). Unambiguous detections of individual
IMBHs do not exist yet, but there are observational hints in favor of
their existence, from studies of ultra-luminous X-ray sources in
nearby star-forming galaxies (Fabbiano 2004; Mushotzky 2004; Miller \& Hamilton 2002).

It has also been long suspected that globular clusters (GCs) may hide IMBHs
in their cores (Shapiro 1977; Shapiro \& Marchant 1978; Marchant \& Shapiro 1979, 1980; Duncan \& Shapiro 1982; see Shapiro 1985 for a review), despite the fact that their formation root is poorly
known: runaway mergers among the most
massive stars, at the time of cluster formation,
can lead to an IMBH, similarly to what
has been conjectured to occur  in young dense star clusters (Portegies Zwart 
et al. 2004; Fregeau et al. 2004).  An alternative pathway is based on
the idea that IMBHs form later in the evolution of GCs
through mergers of stellar-mass BHs. Segregating by dynamical friction
in the core, these BHs are captured in binaries that form 
through dynamical encounters
with stars and BHs. The picture is that hardening by subsequent
interactions with BHs lead them to merge, emitting
gravitational waves (Miller \& Hamilton 2002).  BHs more massive than
stellar-mass BHs are thus created. However, the interactions that
produce hardening also provide recoil, causing the ejection of BHs
(Kulkarni, Hut \& McMillan 1993; Sigurdsson \& Hernquist 1993;
Portegies Zwart \& McMillan 2000), but the heaviest black 
holes may remain, if their mass
exceeds 
a  still uncertain value between 
$\sim 50\msun$ and $\sim 300 \msun$ (Miller \& Hamilton 2002; Colpi, Mapelli \&
Possenti 2003; Gultekin, Miller \& Hamilton 2004).

At present, optical observations of GCs hint in
favor of the existence of IMBHs.  In particular Gebhardt, Rich \& Ho
(2002) suggest the presence of a $2^{+1.4}_{-0.8}
\times{}10^4\,{}M_\odot{}$ IMBH, to explain the kinematics and the
surface brightness profile of the globular cluster G1 in M31. Gerssen
et al. follow the same method to indicate the possible presence of a
$1.7^{+2.7}_{-1.7}\times{}10^3\,{}M_\odot{}$ IMBH in the galactic
globular cluster M15 (Gerssen et al. 2002, 2003). These indications are still controversial, as Baumgardt et al (2003a, 2003b) showed that both the measurements of M15 and those of G1 can be explained with a central concentration of compact objects (neutron stars and white dwarfs) instead of the presence of a massive black hole. However, a more recent paper by Gebhardt, Rich \& Ho (2005) further supports the hypothesis of an IMBH in G1.
 In addition, there has been some claim of the existence of
rotation in the core of a few GCs (Gebhardt et al. 1995; Gebhardt et a. 1997; Gebhardt et al. 2000) suggesting the presence of a steady source of angular
momentum in their cores.

Radio observations can also help in discovering concentrations of
under-luminous matter in the core of GCs, thanks to the presence of
millisecond pulsars that can probe the underlying gravitational field
of the cluster. Currently, there are more than 100 known millisecond
pulsars in GCs (Possenti 2003; Ransom et al. 2005; Camilo \& Rasio
2005) and some show large and negative period derivatives.  This is an
important peculiarity, because millisecond pulsars spin down
intrinsically due to magnetic braking.  Negative period derivatives
come from the variable Doppler shift caused by the acceleration of the
pulsar in the gravitational field of the cluster itself (Phinney
1993).  When these negative period derivatives can be ascribed solely
to the gravitational field of the cluster, these pulsars place a lower
limit on the mass enclosed inside their projected distance from the
globular cluster center.

Two pulsars in M15, at 1'' from the cluster center were found
(Phinney 1993) with negative values of their period derivatives, 
consistent with the mass distribution implied by the stellar
kinematics.  The recent discovery by D'Amico et al. (2002) of two
accelerated pulsars in NGC~6752 at 6'' and 7'' from the center has
highlighted the presence of about $2\times{}10^3 \msun$ of
under-luminous matter in the core of the cluster (Ferraro et
al. 2003a), making NGC~6752 a special target for the search of an IMBH.
NGC~6752 is interesting also because it hosts two pulsars in its halo.
PSR-A is a millisecond pulsar in a binary system
with a white dwarf (Bassa et al. 2003; Ferraro et al. 2003b), located at 3.3 
half mass
radii away: it is the farthest pulsar ever observed in a cluster.
Colpi, Possenti \& Gualandris (2002), and Colpi, Mapelli \& Possenti
(2003, 2004) have suggested that this pulsar has been propelled in the
halo due to a dynamical interaction with a binary IMBH. In modeling
the gravitational encounter, Colpi, Mapelli \& Possenti (2003) found
that a $(50-200,10)\,{}\msun$ binary IMBH is the preferred target for
imprinting the large kick to the pulsar.
Colpi et al. (2005) shortly reviewed the dynamical effects that a
binary  IMBH would imprint on cluster stars.

Considering all the recent hints provided by optical and
radio observations,  we explore, 
in this paper, an new way to unveil a binary IMBH in 
a GC, previously overlooked, that exploits the dynamical 
fingerprint left by a binary IMBH on cluster stars.
In particular we  like to address a number of questions: 
(i) What signature does a binary IMBH imprint on
cluster stars ? Are the stars heated during the scattering process 
still remaining bound to the cluster ?
(ii) Is there direct transfer
of angular momentum from the binary IMBH to the stars to produce some
degree of alignment, given the large inertia of the
BHs ?  (iii) Are prograde or retrograde orbits equally
scattered ?

In this paper, we simulate 3-body encounters of cluster stars with a
binary IMBH, studying the energy and angular momentum exchange, and
reconstructing the trajectories of those stars that are scattered away
from equilibrium by the binary IMBH.  In Section 2 we outline the
method used to simulate 3-body encounters between a binary IMBH and
the cluster stars.  In Section 3 we present our main results: the
formation, in a GC 
 hosting a binary IMBH, of a
population of high velocity (bound) stars (that we call
''supra-thermal stars''), which tend to align their orbital angular
momentum with that of the binary IMBH. In Section 4 we investigate on
the expected number of these supra-thermal stars. In particular, we use an
upgraded version of the code presented in Sigurdsson \& Phinney (1995)
to follow the dynamical evolution of these supra-thermal stars inside
a GC model which reproduces the characteristics of NGC~6752. This
study allows us to put some constraint on the detectability of this family
of high velocity stars. Section 5 contains our conclusions.


\section{The simulations}

We simulated 3-body encounters involving a binary IMBH (composed of
two BHs of mass $M_1$ and $M_2$) and a cluster star of mass
$m=0.5\,{}\msun.$ The dynamics of the encounter is followed solving
the equations of motion 
with a numerical code based on a Runge-Kutta fourth
order integration scheme with adaptive stepsize and quality control
(explained in Colpi, Mapelli \& Possenti 2003, hereafter CMP).  
For the target binary IMBH we considered an interval of
masses between $60\,{}\msun$ up to $210\,{}\msun$. These are the favored
masses of the hypothetical binary IMBH in NGC~6752.  The binary has
semi-major axis $a$ of 1, 10, 100, and 1000 AU and eccentricity\footnote{ We
have repeated select simulations for different values of the
eccentricity and find only minor quantitative differences ($\lesssim{}$10\%) in our results.} $e=0.7$, which is nearly the average binary eccentricity in statistical equilibrium (Hills 1975). The set of models is described in Table~1.

The initial conditions, that are Monte Carlo generated as in CMP, are
sampled using the prescriptions indicated in Hut \& Bahcall (1983).
In particular, the relative velocity
$u^{in}$ is distributed homogeneously between $8.5-11$ km s$^{-1}$
(according to the value of stellar velocity dispersion measured in
NGC~6752 which is our reference cluster; Dubath, Meylan \& Mayor
1997), and the impact parameter $b$ is drawn at random from a
probability distribution uniform in $b^2$ and in a range going from 0 to a truncation value $b_{max}$. The truncation value $b_{max}$ is chosen by requiring that the simulations include all the encounters with non-negligible energetic exchange, i.e. with outgoing velocities of the interacting star significantly higher than the initial velocity  (see Appendix A for 
a discussion of our choices of the impact parameter ranges). 
 The three orientation angles and the phase of the binary are 
generated as indicated in Table~1 and 2 of Hut \& Bahcall (1983). 
We initiate (terminate) integration when the distance between the
incoming (outcoming) star and the center of mass of the binary is
comparable to the radius of gravitational influence of the IMBHs
($r_a\sim{}2G(M_1+M_2)/\sigma{}^2$, where $\sigma$ is the 1-D stellar
dispersion velocity). At the start of each simulation, we place the binary BH and the single star on their respective
hyperbolic trajectories. 

At the end of every scattering experiment we store the final binding
energy $E_{\rm BH}^{fin}$ and angular momentum $\vecJBH^{fin}$ of the
binary IMBH, together with the velocity at infinity (or post-encounter asymptotic velocity) $u^{fin}$, defined as
the velocity of the interacting star after the encounter, estrapolated 
at infinity,
 and the angular
momentum $\vecJstar^{fin}$ of the interacting star. In this paper we
want, in particular, to quantify the mean change in the absolute value
$\Jstar$ of the vector $\vecJstar$ and the extent of alignment of
$\vecJstar$ in the direction of $\vecJBH$.


\begin{table*}
\begin{center}
\caption{Initial Parameters.
}
\begin{tabular}{llllll}
\hline\hline
 & $M_1^{\rm a}$ 
& $M_2^{\rm a}$
& Semi-major axis ($a$)$^{\rm b}$
& Range of impact parameters$^{\rm b}$
& Number of simulations\\
\hline
CASE A1 & 50 & 10 & 1    & 0-60   & 5000\\
CASE A2 & 50 & 10 & 10   & 0-60   & 10000\\
CASE A3 & 50 & 10 & 100  & 0-100  & 10000\\
CASE A4 & 50 & 10 & 1000 & 0-1000 & 10000\\
\noalign{\vspace{0.1cm}}
CASE B1 & 50 & 50 & 1    & 0-100   & 5000\\
CASE B2 & 50 & 50 & 10   & 0-100   & 5000\\
CASE B3 & 50 & 50 & 100  & 0-100   & 5000\\
CASE B4 & 50 & 50 & 1000 & 0-1000  & 5000\\
\noalign{\vspace{0.1cm}}
CASE C1 & 100 & 10 & 1    & 0-100   & 5000\\
CASE C2 & 100 & 10 & 10   & 0-100   & 5000\\
CASE C3 & 100 & 10 & 100  & 0-100   & 5000\\
CASE C4 & 100 & 10 & 1000 & 0-1000  & 5000\\
\noalign{\vspace{0.1cm}}
CASE D1 & 100 & 50 & 1    & 0-100   & 5000\\
CASE D2 & 100 & 50 & 10   & 0-100   & 5000\\
CASE D3 & 100 & 50 & 100  & 0-100   & 5000\\
CASE D4 & 100 & 50 & 1000 & 0-1000  & 5000\\
\noalign{\vspace{0.1cm}}
CASE E1 & 200 & 10 & 1    & 0-100   & 5000\\
CASE E2 & 200 & 10 & 10   & 0-100   & 5000\\
CASE E3 & 200 & 10 & 100  & 0-100   & 5000\\
CASE E4 & 200 & 10 & 1000 & 0-1000  & 5000\\
\noalign{\vspace{0.1cm}}
\hline
\end{tabular}
\begin{flushleft}
{\footnotesize $^{a}$ In units of the solar mass.\\
$^{\rm b}$ In Astronomical Units.}
\end{flushleft}
\end{center}
\end{table*}
\section{Results}
\subsection{Supra-thermal stars}

We find that a sizable number of high velocity stars is created, under
certain conditions, that may highlight the presence of an IMBH in the
cluster.  These stars gain kinetic energy and an excess velocity
relative to the mean, remaining bound to the cluster.  In Table 3 we
give the fraction of bound stars, defined as those having a final
velocity lower than the escape velocity from the core $\sim{}$35-40 km
s$^{-1}$, and the fraction of high velocity or {\it supra-thermal}
stars, defined as having a post-encounter asymptotic (or ''at infinity'') 
speed between 20 and 40
$\kms$. Depending on the characteristics of the binary IMBH, these stars can
account for $\lesssim 50\%$ of all stars that have experienced an
encounter.

Figure 1 shows the distribution of the velocity at infinity 
of stars  scattering off a binary IMBH, 
for cases D1, D2, D3 and D4 (in Table 1).  The shaded area 
indicates the strip of supra-thermal stars.  The mean values of the
speed at infinity $u^{fin}$ and of the fractional binding energy
exchange $\langle{}\Delta{}E_{\rm {BH}}/E_{\rm {BH}}^{in}\rangle{}=\langle{}(E_{\rm {BH}}^{fin}-E_{\rm {BH}}^{in})/E_{\rm {BH}}^{in}\rangle{}$ are given in Table 2
together with the dimensionless factor $\xi{}_E$, defined by:
\begin{equation}
\langle{}\Delta{}E_{\rm {BH}}/E_{\rm{BH}}^{in}\rangle{}=\xi{}_{E}\left [ m/(M_1+M_2)\right] \,{}.
\end{equation}
As illustrated in Figure 1, dynamical encounters widen the velocity
distribution of the stars. This effect is particularly severe
when the binary is hard, having a higher binding energy which becomes 
available in the interaction.
$\xi_E$ clusters around values between 0.6 and 4, and shows a
maximum declining as the binary becomes extremely hard. This explains why case D1 has
a peak velocity at infinity smaller than case D2 corresponding to
a less hard binary (see Table 2). 
We tried to fit the behavior of $\xi_E$, as a function of the 
orbital separation $a$, with a parabola. 
We obtained $\xi_E{}=b\,{}\textrm{Log}^2(a)+c\,{}\textrm{Log}(a)+d$ 
for $b$=-1.2495, $c$=3.8615, $d$=0.958;  this fit is only an approximation, 
since we have few numerical points (Fig. 2). 
 It is interesting to notice that for $a=1$ AU $\xi_E$ is lower than for wider binaries. This does not represent a violation of the first statement of Heggie's law (i.e. that hard binaries tend to become more energetic, Heggie 1975; Hills 1990); in fact all our hard binaries become harder and harder. But a $\xi_E$ decreasing for $a<10$ AU seems in partial disagreement with the second statement of  Heggie's law, for which the hardening rate of the binaries is ''approximately independent of their binding energy'' (Heggie 1975). This difference is the result of a different approach with respect to that followed by Heggie. In fact, we are treating very massive binaries with respect to the incoming star, and such binaries require very large maximum impact parameters, nearly independently of their semi-major axis, to include in our simulations all the interactions which have significant energy exchange (i.e. $\Delta{}E_{{\rm BH}}/E^{in}_{{\rm BH}}>10^{-3}$), corresponding to post-encounter asymptotic velocities of the star in the supra-thermal range (see Appendix A for a complete discussion).

In Figure 1, case D1 and D3 give the highest number of supra-thermal
stars among the different runs.  These supra-thermal stars are the
ones that have also acquired a sizable fraction of the orbital
angular momentum of the binary IMBH. Thus in scattering off the binary
IMBH they are preferentially launched in a halo orbit, inside the
cluster.  Their detection (discussed in $\S$ 4) would be the sign of
an IMBH hidden in the cluster.

\begin{figure}
\center{{
\epsfig{figure=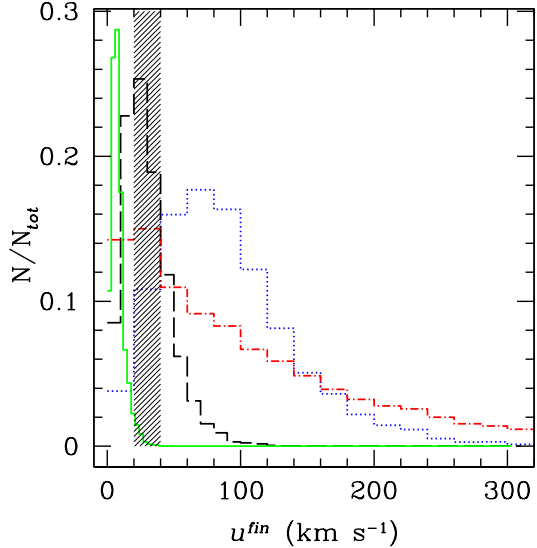,height=8cm}
}}
\caption{\label{fig:fig1} 
Post-encounter asymptotic velocity distributions of the
cluster star in the case D1 ({\it dotted-dashed line}), D2 ({\it
dotted line}), D3 ({\it dashed line}), D4 ({\it solid line}). The
shaded area refers to the supra-thermal stars.  On the y-axis the
number of cases for each bin is normalized to the total number of
resolved runs (N$_{tot}$).}
\end{figure}


\begin{figure}
\center{{
\epsfig{figure=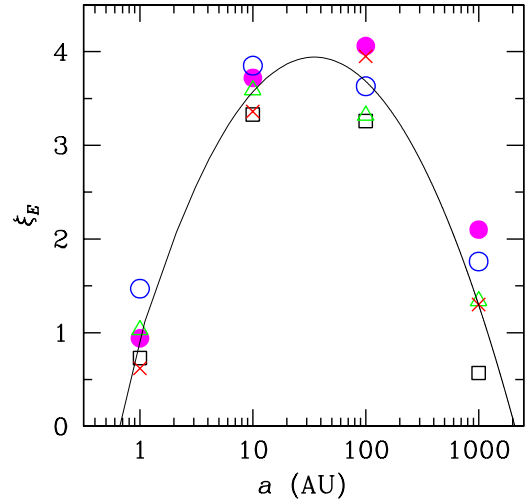,height=8cm}
}}
\caption{\label{fig:fig2} 
$\xi{}_E\equiv{}[(M_1+M_2)/m]\langle{}\Delta{}E_{{\rm BH}}/E^{in}_{{\rm BH}}\rangle{}$
as a function of the orbital separation $a$ for all the considered
cases. A1, A2, A3 and A4 are represented by {\it open triangles}; B1,
B2, B3 and B4 by {\it crosses}; C1, C2, C3 and C4 by {\it open
squares}; D1, D2, D3 and D4 by {\it filled circles}; E1, E2, E3 and E4
by {\it open circles}. The {\it solid line} indicates a parabolic fit
$b\,{}\textrm{Log}^2(a)+c\,{}\textrm{Log}(a)+d=\xi{}_E(a)$, where
$b$=-1.2495, $c$=3.8615, $d$=0.958.  }
\end{figure}


\begin{figure}
\center{{
\epsfig{figure=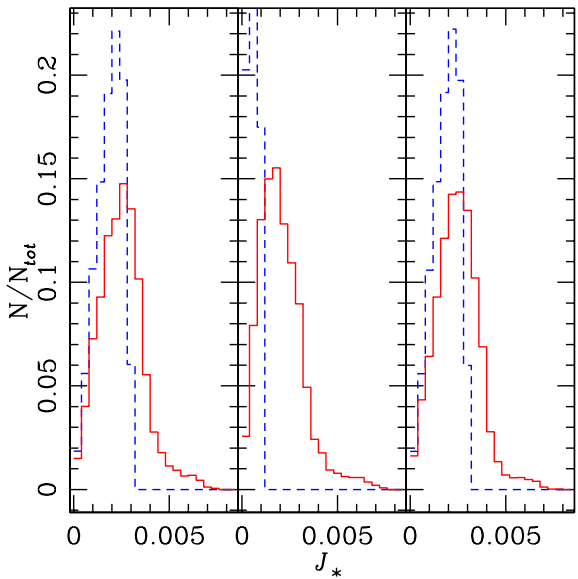,height=8cm}
}}
\caption{\label{fig:fig3} 
Angular momentum distribution of the cluster
star in the case D2 ({\it left panel}), D3 ({\it central panel}), D4
({\it right panel}). The angular momenta are in dimensionless units
(i.e. normalized to the length, mass and time scales). {\it Dashed
line} indicates the initial angular momentum distribution, {\it solid
line} indicates the post-encounter angular momentum distribution. 
Evidence of angular momentum transfer from the binary to the cluster star
occurs when the binary has an semi-major axis of 100 AU (case D3);
while the softest binary (case D4, $a$=1000 AU) is the least
efficient from this point of view.  On the y-axis the number of cases for each bin is normalized to the total number of resolved runs (N$_{tot}$).}
\end{figure}




\begin{table*}
\begin{center}
\caption{Final velocities and angular momentum exchanges.$^{\rm a}$}  
\begin{tabular}{llllll}
\hline
\hline
 & $u^{fin\quad{}\rm b}$
&  $\langle{}\frac{\Delta{}J}{J_{\ast{}}^{in}}\rangle{}^{\rm c}$
 & $\frac{\Delta{}J}{J_{\ast{}}^{in}}^{\rm d}$
&  $\langle{}\frac{\Delta{}E_{\rm BH}}{E_{\rm{BH}}^{in}}\rangle{}^{\rm e}$
&  $\xi{}_E^{\rm f}$\\
\hline
CASE A1 
& 12.5$_{-10.0}^{+95.0}$ & 0.056$_{-0.054}^{+14.40}$ & 0.025$_{-0.100}^{+0.100}$ 
& 0.0086$_{-0.0073}^{+0.4520}$ & 1.03\\
\noalign{\vspace{0.05cm}}
CASE A2 
& 37.5$_{-30.0}^{+35.0}$ & 0.630$_{-0.592}^{+8.24}$ & 0.175$_{-0.850}^{+1.150}$ 
& 0.0299$_{-0.0186}^{+1.1032}$ & 3.59\\
\noalign{\vspace{0.05cm}}
CASE A3 
& 12.5$_{-10.0}^{+10.0}$ & 1.12$_{-1.13}^{+1.82}$ & 0.725$_{-1.400}^{+2.200}$    
& 0.0277$_{-0.0191}^{+1.0363}$ & 3.32\\
\noalign{\vspace{0.05cm}}
CASE A4 
& 7.5$_{-3.0}^{+2.0}$  &  0.155$_{-0.228}^{+30.330}$ &  0.025$_{-0.500}^{+0.350}$ 
& 0.0112$_{-0.0199}^{+0.0309}$ & 1.34\\
\noalign{\vspace{0.1cm}}
CASE B1 
& 12.5$_{-12.5}^{+75.0}$  & 0.037$_{-0.037}^{+7.980}$ & 0.005$_{-0.005}^{+0.020}$ 
&  0.0031$_{-0.0030}^{+0.1778}$ & 0.62\\
\noalign{\vspace{0.05cm}}
CASE B2 
& 67.5$_{-55.0}^{+60.0}$  & 0.317$_{-0.346}^{+31.840}$ & 0.025$_{-0.55}^{+0.65}$ 
&  0.0168$_{-0.0109}^{+0.2621}$ & 3.36\\
\noalign{\vspace{0.05cm}}
CASE B3 
& 17.5$_{-15.0}^{+25.0}$  &  1.994$_{-1.456}^{+145.935}$ &  0.325$_{-1.100}^{+2.850}$ 
& 0.0198$_{-0.0134}^{+0.2255}$ & 3.95\\
\noalign{\vspace{0.05cm}}
CASE B4 
& 8.5$_{-6.0}^{+3.0}$  &  0.197$_{-0.259}^{+17.279}$ & -0.0255$_{-0.400}^{+0.550}$ 
& 0.0065$_{-0.0073}^{+0.0150}$ & 1.30\\
\noalign{\vspace{0.1cm}}
CASE C1 
& 12.5$_{-10.0}^{+60.0}$  & 0.020$_{-0.020}^{+5.726}$ & 0.005$_{-0.005}^{+0.010}$ 
& 0.0033$_{-0.0032}^{+0.7152}$ & 0.73 \\
\noalign{\vspace{0.05cm}}
CASE C2 
& 32.5$_{-25.0}^{+40.0}$  & 0.154$_{-0.284}^{+0.891}$ & 0.075$_{-0.550}^{+0.550}$ 
 & 0.0151$_{-0.0095}^{+0.4422}$ & 3.33 \\
\noalign{\vspace{0.05cm}}
CASE C3 
& 12.5$_{-10.0}^{+10.0}$ &  0.876$_{-1.135}^{+1.639}$ & 0.520$_{-1.040}^{+3.200}$ 
& 0.0148$_{-0.0098}^{+0.5524}$ & 3.26\\
\noalign{\vspace{0.05cm}}
CASE C4 
& 5.9$_{-3.1}^{+2.5}$ &  0.212$_{-0.255}^{+21.888}$ & 0.025$_{-0.500}^{+0.500}$ 
& 0.00261$_{-0.00773}^{+0.00911}$ & 0.57\\
\noalign{\vspace{0.1cm}}
CASE D1 
& 30.0$_{-30.0}^{+160.0}$ & 0.062$_{-0.059}^{+12.175}$ & 0.005$_{-0.050}^{+0.090}$ 
& 0.0031$_{-0.0027}^{+0.1054}$ & 0.94\\
\noalign{\vspace{0.05cm}}
CASE D2 
& 75.0$_{-60.0}^{+70.0}$ & 0.205$_{-0.623}^{+0.703}$ & 0.225$_{-1.000}^{+0.850}$ 
& 0.0124$_{-0.0078}^{+0.1679}$ & 3.72\\
\noalign{\vspace{0.05cm}}
CASE D3 
& 22.5$_{-20.0}^{+30.0}$ & 2.911$_{-1.801}^{+7.894}$ & 1.235$_{-1.890}^{+3.780}$
& 0.0135$_{-0.0090}^{+0.0348}$ & 4.06 \\
\noalign{\vspace{0.05cm}}
CASE D4 
   & 7.0$_{-6.0}^{+6.0}$  & 0.384$_{-0.512}^{+1.243}$ & 0.185$_{-0.840}^{+1.050}$ 
& 0.0070$_{-0.0065}^{+0.7796}$ & 2.10\\
\noalign{\vspace{0.1cm}}
CASE E1 
& 27.5$_{-25.0}^{+105.0}$ &  0.025$_{-0.024}^{+7.087}$ & 0.005$_{-0.030}^{+0.050}$ 
& 0.0035$_{-0.0027}^{+0.7193}$ & 1.47\\
\noalign{\vspace{0.05cm}}
CASE E2 
& 32.5$_{-25.0}^{+45.0}$ &  0.429$_{-0.381}^{+28.250}$ & 0.065$_{-0.420}^{+0.540}$ 
& 0.0092$_{-0.0058}^{+0.6392}$ & 3.85\\
\noalign{\vspace{0.05cm}}
CASE E3 
& 13.5$_{-12.0}^{+12.0}$  & 1.212$_{-0.881}^{+3.718}$ & 0.675$_{-1.700}^{+1.350}$ 
& 0.0086$_{-0.0060}^{+0.6697}$ & 3.63\\
\noalign{\vspace{0.05cm}}
CASE E4 
& 6.5$_{-5.0}^{+3.0}$ & 0.395$_{-0.347}^{+50.771}$ & 0.075$_{-0.400}^{+0.450}$ 
& 0.0042$_{-0.0067}^{+0.0191}$ & 1.76\\
\noalign{\vspace{0.05cm}}
\hline
\end{tabular}
\begin{flushleft}
{\footnotesize $^{\rm a}$
We consider both bound and ejected stars. Only unresolved encounters are neglected in this Table.\\
$^{\rm b}$ In units of km s$^{-1}$. Peak value of the velocity at infinity of the star cluster. The dispersion around the peak value is 
calculated considering those values which contain $50\%{}$ of the
total area  descending from the peak.\\
$^{\rm c}$ $\Delta{}J\equiv{}(J_{\ast{}}^{fin}-J_{\ast{}}^{in})$, where $J_{\ast{}}^{in}$ and $J_{\ast{}}^{fin}$ represent respectively the modulus of the initial and the final angular momentum of the cluster star. $\langle{}\frac{\Delta{}J}{J_{\ast{}}^{in}}\rangle{}$ is the mean value of the variation of the absolute value of the angular momentum of the cluster star, normalized to its initial value $J_{\ast{}}^{in}$. The dispersion around the mean value is calculated considering those values which contain $34\%{}$ of the total area in the left and  right wings, respectively.\\
$^{\rm d}$ $\frac{\Delta{}J}{J_{\ast{}}^{in}}$ is the peak value of $\frac{\Delta{}J}{J_{\ast{}}^{in}}$.  The dispersion around the peak value is 
calculated considering those values which contain $50\%{}$ of the
total area  descending from the peak.\\
$^{\rm e}$ $\langle{}\frac{\Delta{}E_{\rm BH}}{E_{\rm{BH}}^{in}}\rangle{}$ is the mean value of the variation of the binding energy of the binary IMBH, normalized to its initial binding energy $E_{\rm{BH}}^{in}$. The dispersion around the mean value is calculated considering those values which contain $34\%{}$ of the total area in the left and  right wings, respectively.\\
$^{\rm f}$ $\xi{}_E$ represents the hardening factor and is given by $\xi{}_E\equiv{}\frac{M_1+M_2}{m}\langle{}\frac{\Delta{}E_{\rm BH}}{E_{\rm{BH}}^{in}}\rangle{}$ (CMP 2003).
 }
\end{flushleft}
\end{center}
\end{table*}



\begin{table*}
\begin{center}
\caption{Statistics of the Outgoing States.  
}
\begin{tabular}{llllll}
\hline
\hline
 & Bound stars (\%)
 & Ejections (\%)
& Unresolved Encounters (\%)$^{\rm a}$
& Supra-thermal stars (\%)$^{\rm b}$\\
\hline
CASE A1 &  45.02 & 53.12 & 1.86 & 21.46\\
CASE A2 & 51.32 & 47.10 & 1.58 & 35.56\\
CASE A3 & 97.60 & 1.65 & 0.75  & 17.23\\
CASE A4 & 99.98 & 0.00 & 0.02  &  0.06\\
\noalign{\vspace{0.1cm}}
CASE B1 & 53.78 & 43.44 & 2.78 & 13.12\\
CASE B2 & 21.12 & 77.90 & 0.98 & 15.58\\
CASE B3 & 83.36 & 16.30 & 0.34 & 43.36\\
CASE B4 & 99.96 &  0.00 & 0.04 &  0.36\\
\noalign{\vspace{0.1cm}}
CASE C1 & 60.46 & 34.82 & 4.72 & 15.10  \\
CASE C2 & 51.30 & 45.58 & 3.12 & 35.24  \\
CASE C3 & 95.56 &  1.80 & 2.64 & 17.22 \\
CASE C4 & 99.86 &  0.00 & 0.14 &  0.01 \\
\noalign{\vspace{0.1cm}}
CASE D1 & 28.04 & 67.84 & 4.12 &  14.38 \\
CASE D2 & 14.44 & 84.18 & 1.38 &  10.68 \\
CASE D3 & 75.20 & 24.32 & 0.48 &  44.04\\
CASE D4 & 99.84 &  0.04 & 0.12 &   3.46\\
\noalign{\vspace{0.1cm}}
CASE E1 & 27.60 & 62.54 & 9.86 &  17.86\\
CASE E2 & 45.84 & 47.42 & 6.74 &  31.54\\
CASE E3 & 88.68 &  2.36 & 8.96 &  18.42\\
CASE E4 & 98.48 & 0.02 & 1.50 & 0.38 \\
\noalign{\vspace{0.1cm}}
\hline
\end{tabular}
\begin{flushleft}
{\footnotesize $^{\rm a}$ We define unresolved interactions those runs where the separation between the star and the binary IMBH never exceeds 30 $a$ after 10$^8$ time steps.\\
$^{\rm b}$ We define supra-thermal stars those with  velocities at infinity  comprised between 20 and 40 km s$^{-1}$.}
\end{flushleft}
\end{center}
\end{table*}


\subsection{Angular Momentum Transfer and Alignment}

An interesting question to address is whether and how the orbital
angular momentum of the binary IMBH couples to the star after a close
gravitational encounter.  The BHs can be sufficiently massive and the
orbit be sufficiently wide that the orbital angular momentum of the
binary exceeds that of the incoming star.  In this case a direct
transfer of orbital angular momentum can occur. The star coming close
to the binary IMBH can be dragged into corotation, i.e. the star can
emerge after the encounter with an angular momentum nearly aligned
with the binary IMBH.

If we denote with $\mu{}=M_1 M_2/(M_1+M_2)$ the reduced mass of the binary hosting the two BHs, the total angular momentum of the system is 

\begin{equation}
\label{eq:eq4}
\begin{array}{l}
{\bf {J}} = \vecJBH{}^{in} + \vecJstar{}^{in}\\
\hspace{0.3cm}= \mu \sqrt{a\,{}G\,(M_1+M_2)}{\bf{z}}
 + 
b\,{}u^{in} \left[\frac{m(M_1+M_2)}{M_1+M_2+m}\right]{\bf {z'}}
\end{array}
\end{equation}
where $\bf{z}$ and $\bf{z}'$ are the unit vectors indicating 
respectively the directions of $\vecJBH^{in}$ and $\vecJstar^{in}$. 
Angular momentum transfer from the binary to
the interacting star and  partial alignment become important
when
$J_{BH}^{in}\gg{}J_*^{in}$, i.e., when
\begin{equation}\label{eq:eq4}
\frac{\mu}{b\,{}u^{in}\,{}m}\sqrt{(M_1+M_2)\,{}a\,{}G}\gg{}1. 
\end{equation}
Our experiment confirms this fact. Figure 3 ({\it solid lines}) shows the 
post-encounter
distribution of $\Jstar$ for cases D2, D3 and D4.  It is compared with the
distribution of $\Jstar$ of the incoming stars ({\it dashed lines}) to show
that the binary IMBH transfers angular momentum to the stars widening
the distribution of $\Jstar$.  Table 2 contains the averaged value of
the fractional angular momentum increase (in modulus) $\langle{}\Delta
\Jstar/\Jstar{}^{in}\rangle{}$ for the complete series of runs.  $\langle{}\Delta
\Jstar/\Jstar^{in}\rangle{}$ is large and exceeds unity in correspondence to
the runs where the bulk of the supra-thermal stars are produced. It
has a non monotonic trend with the hardness 
of the binary: a harder
binary has little excess of angular momentum in its initial state
relative to that of the incoming star; so the efficiency of angular
momentum transfer reduces.  A less hard binary (having a larger
$J_{\rm BH}^{in}$) is also less efficient, since it does not alter much
the post-encounter asymptotic velocity.  

We can further quantify
the importance  of angular momentum transfer from the binary to the
star, by computing the fraction of stars which increase $\Jstar$ 
after an interaction; this is given in the last column
of Table 4.  The fraction of stars that do so, over the total (in the
sample of the bound stars), is significantly high: it is above 70$\%$
in many cases.


\begin{table*}
\begin{center}
\caption{Statistics for bound stars$^{\rm a}$.}
\begin{tabular}{lllll}
\hline
\hline
& Corot$_{in}$ (\%)$^{\rm b}$ 
& Corot$_{fin}$ (\%)$^{\rm c}$
& More aligned stars (\%)$^{\rm d}$
& $J_{\ast{}}^{fin}>J_{\ast{}}^{in}$ (\%)$^{\rm e}$\\
\hline
CASE A1 & 40 & 42 & 52 & 64 \\
CASE A2 & 47 & 59 & 61 & 72 \\
CASE A3 & 49 & 68 & 65 & 87 \\
CASE A4 & 50 & 51 & 50 & 55 \\
\noalign{\vspace{0.1cm}}
CASE B1 & 46 & 46 & 50 & 59\\
CASE B2 & 57 & 63 & 53 & 64\\
CASE B3 & 50 & 58 & 56 & 83\\
CASE B4 & 50 & 58 & 55 & 58\\
\noalign{\vspace{0.1cm}}
CASE C1 & 46 & 46 & 51 & 59\\ 
CASE C2 & 45 & 49 & 58 & 64\\ 
CASE C3 & 49 & 62 & 64 & 88\\
CASE C4 & 50 & 52 & 51 & 59\\
\noalign{\vspace{0.1cm}}
CASE D1 & 40 & 40 & 51 & 64\\
CASE D2 & 55 & 60 & 52 & 71\\
CASE D3 & 51 & 70 & 62 & 95\\
CASE D4 & 50 & 69 & 63 & 71\\
\noalign{\vspace{0.1cm}}
CASE E1 & 40 & 40 & 50 & 61\\
CASE E2 & 42 & 45 & 60 & 65\\
CASE E3 & 48 & 53 & 61 & 92 \\
CASE E4 & 50 & 54 & 59 & 63 \\
\noalign{\vspace{0.1cm}}
\hline
\end{tabular}
\begin{flushleft}
{\footnotesize
$^{\rm a}$ In this Table we consider only the stars which after the interaction with the binary remain bound to the cluster.\\
$^{\rm b}$ Percentage (respect to the total of bound stars) of stars which, before the interaction, are corotating with the binary (i.e. for which the scalar product between their angular momentum and the angular momentum of the binary is positive). \\
$^{\rm c}$ Percentage (respect to the total of bound stars) of stars which, after the interaction, are corotating with the binary, independently from the initial orientation of their angular momentum.\\
$^{\rm d}$ Percentage of bound stars which, after the encounter, reduce the angle between their angular momentum and that of the binary.\\
$^{\rm e}$ Percentage of bound stars which, after the encounter, increase the absolute value of their angular momentum.
}
\end{flushleft}
\end{center}
\end{table*}


Is alignment induced in the scattering process ?  In Table 4 we
compare the percentage of bound stars which were corotating (i.e. for
which the scalar product between their angular momentum and the
angular momentum of the binary is positive, that is
$\vecJstar\cdot{}\vecJBH>0$) before the 3-body interaction with the
percentage of those stars which are corotating after the 3-body
interaction. Whereas the fraction of corotating stars before the
interaction is $\sim{}$50\% (as one can expect given the initial
sampling of the data), the fraction of corotating stars after the
interaction is often over 60\%, indicating a tendency of stars to
align their angular momentum with that of the binary IMBH. This
tendency is greater if the binary has a big reduced mass (cases with
$M_1=100,\,{}M_2=50\,{}M_\odot{}$ and with
$M_1=50,\,{}M_2=50\,{}M_\odot{}$) and if the binary is moderately hard
(i.e. its orbital separation is neither too small, because in this
case $J_{\rm {BH}}^{in}$ is comparable with $J_{\ast{}}^{in}$, nor too
large, since in this case the binary would be too soft, and so the
kinetic energy exchange would be negligible).  In case D3 the bound
stars which after the interaction are corotating with the binary are
$\sim{}$70\%. This means that we can observe a family of supra-thermal
stars which have an anisotropy in their orbital angular momentum
relative to the center of the cluster, 70\% of them rotating in one
direction and the remaining 30\% rotating in the other. Obviously, we
have to take into account that orbits initially are isotropically
distributed and that we observe them in projection; then the detection
of this phenomenon is not so immediate.

 Table 4 indirectly shows also another interesting effect. The first column indicates that generally less than 50\% of the initially corotating stars remain bound to the cluster, while the post-encounter percentage of corotating stars is more than 50\%. This means not only that a fraction of initially counter-rotating stars becomes corotating, but also that an initially corotating star is more easy ejected from the cluster than a counter-rotating star. This fact can be intuitively explained considering that, when  a counter-rotating star interacts with the binary, its relative velocity with respect to the lighter BH is higher than in the case of a corotating star, and, then, the cross section is lower.

To better constrain orbit alignment in the direction of $\vecJBH$,
we considered separately
orbits which before the encounter (as initial condition) were
corotating (Table 5) 
and orbits which before the encounter were
counterrotating (Table 6). For both we investigated the final distributions, 
and in particular we compared the fraction of stars initially
corotating which after the interaction remain corotating (Table 5, 2nd
column) with the fraction of stars initially
counterrotating which after the interaction become corotating (Table 6,
2nd column). We found that most (at least 70\%) of the initially
corotating stars remains corotating, even if during very energetic
interactions -with the hardest binaries- the star tends to forget its
initial angular momentum orientation (case D2). In contrast, a high
percentage of initially counterrotating stars becomes corotating,
flipping their angular momentum. The combination of these two tendencies
(i.e. the tendency to remain corotating for initially corotating stars
and the tendency to become corotating for initially counterrotating
stars) provides the strongest evidence of an alignment between
$\vecJBH^{in}$ and $\vecJstar^{in}$ after the interaction (Fig. 4).In Fig. 5 we
give a complete overview of the cases considered. 


\begin{table*}
\begin{center}
\caption{Statistics for initially corotating bound stars$^{\rm a}$.}
\begin{tabular}{lllll}
\hline
\hline
 & Corot$_{fin}$ (\%)$^{\rm b}$ 
 & Counterrot$_{fin}$ (\%)$^{\rm c}$ 
 & More aligned stars (\%)$^{\rm d}$
 & $J_{\ast{}}^{fin}>J_{\ast{}}^{in}$ (\%)$^{\rm e}$\\
\hline
CASE A1 & 93 & 7 & 49 & 70\\
CASE A2 & 77 & 23 & 44 & 77\\
CASE A3 & 79 & 21 & 48 & 90\\
CASE A4 & 82 & 18 & 42 & 53\\
\noalign{\vspace{0.1cm}}
CASE B1 & 97 &  3 & 50 & 63\\
CASE B2 & 67 & 33 & 34 & 67\\
CASE B3 & 60 & 40 & 35 & 84\\
CASE B4 & 62 & 38 & 33 & 61\\
\noalign{\vspace{0.1cm}}
CASE C1 & 98 &  2 & 50 & 62\\ 
CASE C2 & 81 & 19 & 46 & 73\\ 
CASE C3 & 86 & 14 & 48 & 91\\
CASE C4 & 86 & 14 & 45 & 60\\
\noalign{\vspace{0.1cm}}
CASE D1 & 92 &  8 & 50 & 68 \\
CASE D2 & 65 & 35 & 31 & 67 \\
CASE D3 & 72 & 28 & 41 & 93 \\
CASE D4 & 75 & 25 & 43 & 71 \\
\noalign{\vspace{0.1cm}}
CASE E1 & 94 &  6 & 48 & 63 \\
CASE E2 & 86 & 14 & 48 & 73 \\
CASE E3 & 96 &  4 & 35 & 95 \\
CASE E4 & 94 &  6 & 53 & 70 \\
\noalign{\vspace{0.1cm}}
\hline
\end{tabular}
\begin{flushleft}
{\footnotesize
$^{\rm a}$ In this Table we consider only the stars which before the interaction were corotating.\\
$^{\rm b}$ Percentage (respect to the total of initially corotating bound stars) of stars which, after the interaction, remain corotating with the binary.\\
$^{\rm c}$ Percentage (respect to the total of initially corotating bound stars) of stars which, after the interaction, become counterrotating respect to the binary.\\
$^{\rm d}$ Percentage (respect to the total of initially corotating bound stars) of stars which, after the encounter, reduce the angle between their angular momentum and that of the binary.\\
$^{\rm e}$ Percentage of bound stars (respect to the total of initially corotating bound stars) which, after the encounter, increase the absolute value of their angular momentum.
}
\end{flushleft}
\end{center}
\end{table*}

\begin{table*}
\begin{center}
\caption{Statistics for initially counterrotating bound stars$^{\rm a}$.
}
\begin{tabular}{lllll}
\hline\hline
 & Corot$_{fin}$ (\%)$^{\rm b}$ 
 & Counterrot$_{fin}$ (\%)$^{\rm c}$
 & More aligned stars (\%)$^{\rm d}$
 & $J_{\ast{}}^{fin}>J_{\ast{}}^{in}$ (\%)$^{\rm e}$\\
\hline
CASE A1 &  7 & 93 & 54 & 60\\
CASE A2 & 43 & 57 & 75 & 69\\
CASE A3 & 57 & 43 & 81 & 85\\
CASE A4 & 21 & 79 & 58 & 57\\
\noalign{\vspace{0.1cm}}
CASE B1 &  3 & 97 & 49 & 57\\
CASE B2 & 57 & 43 & 79 & 60\\
CASE B3 & 56 & 44 & 77 & 82\\
CASE B4 & 55 & 45 & 79 & 55\\
\noalign{\vspace{0.1cm}}
CASE C1 & 34 & 66 & 52 & 57\\ 
CASE C2 & 23 & 76 & 68 & 57\\ 
CASE C3 & 39 & 61 & 79 & 84 \\
CASE C4 & 17 & 83 & 57 & 58 \\
\noalign{\vspace{0.1cm}}
CASE D1 &  6 & 94 & 52 & 61 \\
CASE D2 & 54 & 46 & 78 & 76 \\
CASE D3 & 67 & 33 & 84 & 97 \\
CASE D4 & 62 & 38 & 83 & 71 \\
\noalign{\vspace{0.1cm}}
CASE E1 &  5 & 95 & 52 & 59  \\
CASE E2 & 15 & 85 & 69 & 59  \\
CASE E3 & 13 & 87 & 85 & 90  \\
CASE E4 & 15 & 85 & 65 & 57 \\
\noalign{\vspace{0.1cm}}
\hline
\end{tabular}
\begin{flushleft}
{\footnotesize
$^{\rm a}$ In this Table we consider only the stars which before the interaction were counterrotating.\\
$^{\rm b}$ Percentage (respect to the total of initially counterrotating bound stars) of stars which, after the interaction, become corotating with the binary.\\
$^{\rm c}$ Percentage (respect to the total of initially counterrotating bound stars) of stars which, after the interaction, remain counterrotating respect to the binary.\\
$^{\rm d}$ Percentage (respect to the total of initially counterrotating bound stars) of stars which, after the encounter, reduce the angle between their angular momentum and that of the binary.\\
$^{\rm e}$ Percentage of bound stars (respect to the total of initially counterrotating bound stars) which, after the encounter, increase the absolute value of their angular momentum.
}
\end{flushleft}
\end{center}
\end{table*}

\begin{figure}
\center{{
\epsfig{figure=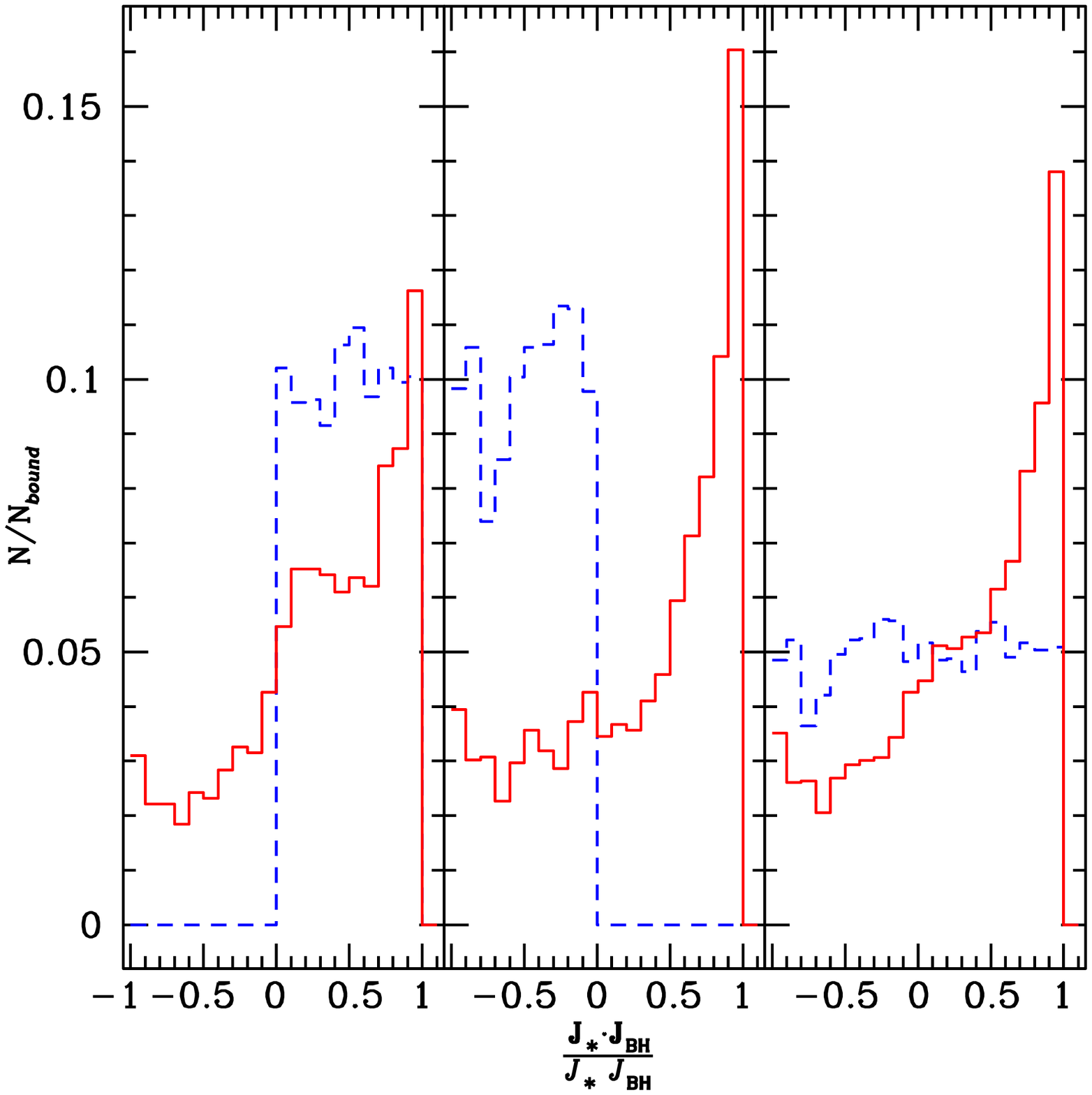,height=8cm}
}}
\caption{\label{fig:fig4} 
The histograms show the distribution of
$\frac{\vecJstar\cdot{}\vecJBH}{J_{\ast{}}\,{}J_{\rm {BH}}}$ for the case D3
($M_1=100\,{}M_\odot{},\,{}M_2=50\,{}M_\odot{},\,{}a=100$ AU). {\it
Solid line} indicates the distribution after the encounter; {\it
dashed line} indicates the distribution before the encounter. The left
panel represents the distribution of
$\frac{\vecJstar\cdot{}\vecJBH}{J_{\ast{}}\,{}J_{\rm {BH}}}$ for bound stars
which before the encounter were corotating, the central panel the
distribution for bound stars which before the encounter where
counterrotating and the right panel the sum of the two distribution,
i.e. the distribution for all the stars which after the encounter
remain bound to the cluster.  On the y-axis the number of cases
for each bin is normalized to the total number of bound stars
(N$_{bound}$).
}
\end{figure}


\begin{figure}
\center{{
\epsfig{figure=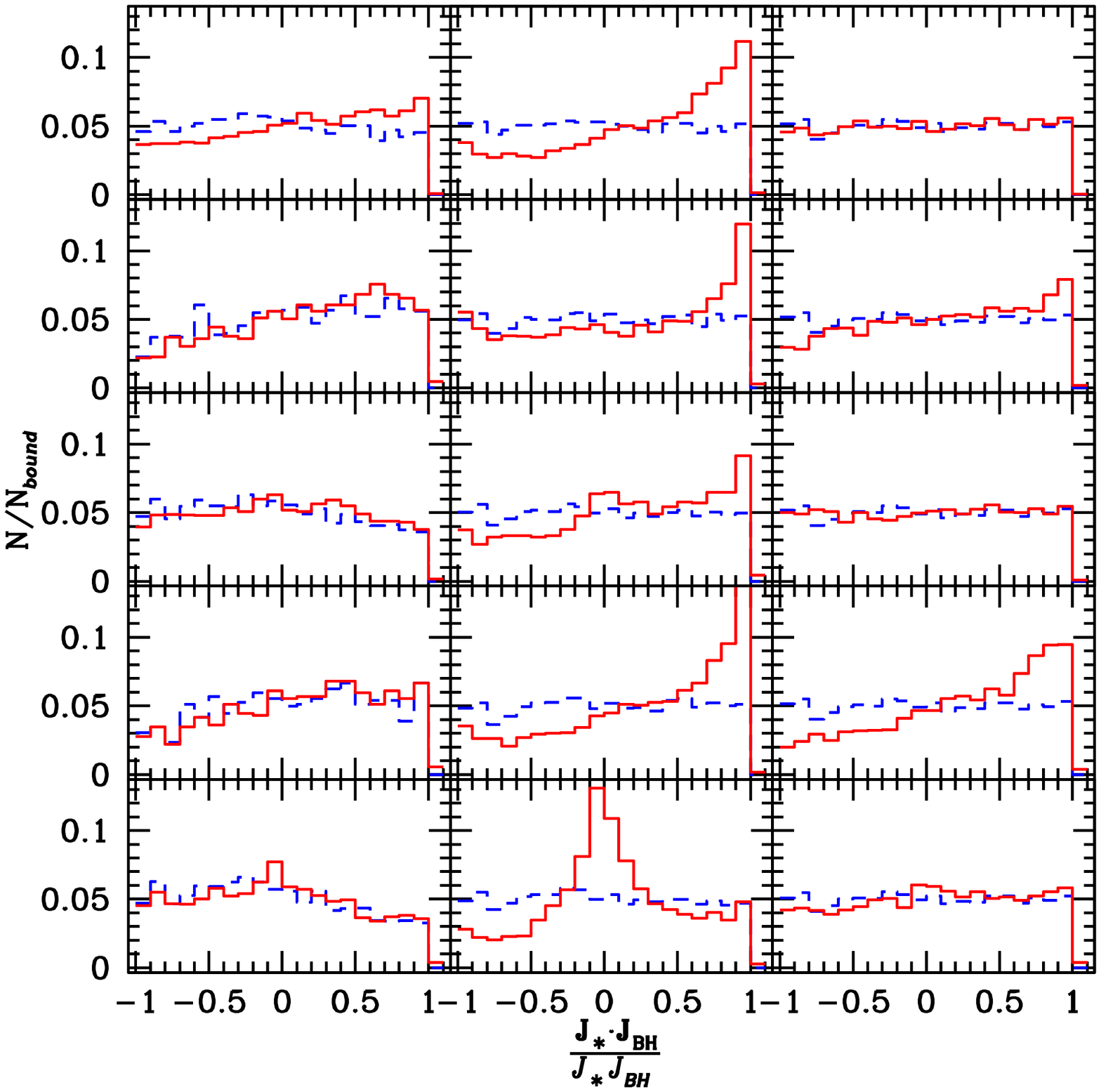,height=8cm}
}}
\caption{\label{fig:fig5} 
The histograms show the distribution of
$\frac{\vecJstar\cdot{}\vecJBH}{J_{\ast{}}\,{}J_{\rm {BH}}}$ for all the
considered cases. {\it Solid line} indicates the distribution after
the encounter; {\it dashed line} indicates the distribution before the
encounter. From the up left panel, going toward right: case A2, A3,
A4, B2, B3, B4, C2, C3, C4, D2, D3, D4, E2, E3, E4.  We do not
plot cases A1, B1, C1, D1 and E1 because they are similar to cases
A2, B2, C2, D2 and E2. On the y-axis the number of cases for each bin
is normalized to the total number of bound stars (N$_{bound}$).  }
\end{figure}


\section{Detecting supra-thermal stars}
We want now to estimate how many supra-thermal stars
are produced by interactions with a binary IMBH and if they 
are observable with present instrumentation.

\subsection {Estimating the number of supra-thermal stars}

Since supra-thermal stars have excess kinetic energy and angular momentum, they are not in thermodynamic
equilibrium with respect to the rest of the cluster. Thus, two-body 
relaxation will try to obliterate such
anisotropies. Friction will return them to the core where they become
unrecognizable, unless some memory of the angular momentum alignment
(e.g. Section 3.2) is retained during the relaxation process.

The existence of supra-thermal stars requires:  
\begin{itemize}
\item[(i)] the survival of the IMBH as
a binary in the cluster core;
\item[(ii)] the persistence of the
supra-thermal stars in the GC halo.
\end{itemize}
We can investigate these questions through a comparison between relevant timescales.
The first timescale to consider is the
hardening time of the binary IMBH (Quinlan 1996), given the stellar number density
in the core $n$ and the binary separation $a$

\begin{equation}\label{eq:eq1}
t_{hard}=\frac{\sigma{}}{2\pi{}a \, \xi{}_E\,{}G\, m\,{}n}
\end{equation}
The second relevant timescale is the relaxation time at the half mass radius
\begin{equation}
t_{rh}=\frac{0.14N}{\ln{(0.4N)}}\left(\frac{r_h^3}{GM}\right)^{1/2}
\end{equation}
where $N$ and $M$ are respectively the number of stars and the total
mass of the cluster.  In Figure 6 we compare the two timescales for
$\sigma=8.5\kms$, $\xi_E\sim 1$ (see Table 2) and $n\,{}=10^5$ stars
pc$^{-3}$.  This plot shows that, when the binary is not very hard,
i.e., when $a>\,{}4$ AU, the hardening time is shorter than the
half-mass relaxation time.  This implies that we can simultaneously
observe all the supra-thermal stars in the halo for a time comparable
to $t_{rh}$ with no dilution.  On the other hand, when the
binary has orbital separation $a\leq{}\,{}4$ AU, $t_{rh}$ is
shorter than $t_{hard}$, and, then, we can not simultaneously
observe all the produced supra-thermal stars in the halo, a part of
them being already thermalized. However, this effect is balanced by
the fact that the binary remains in this very hard configuration for
most of its life, and, therefore, we can observe a considerable
fraction ($\sim{}60$\%) of the total number of supra-thermal stars
even when the binary is very hard ($a\leq{}\,{}4$ AU).\\ For
completeness, we also considered a third timescale: the gravitational wave
timescale, $t_{gw}$, i.e. the characteristic time 
for a binary IMBH to coalesce due to gravitational wave
emission.  When this timescale,  defined as
(Peters 1964; Quinlan 1996)
\begin{equation}\label{eq:tgw}
t_{gw}=\frac{5}{256}\frac{c^{5}a^{4}(1-e^2)^{7/2}}{G^{3}M_1M_2(M_1+M_2)},
\end{equation}
becomes shorter than $t_{hard}$, the binary would  
shrink, due to gravitational wave emission, at a rate shorter than what we 
expected from $t_{hard}$. 
Figure~6 shows the timescale $t_{gw}$ computed for
a constant eccentricity $e=0.7$ (the initial eccentricity we chose for our runs) and for the range
of masses considered in this paper. We expect that the assumption of constant eccentricity is correct for star-binary IMBH interactions, since, when $M_1+M_2\gg{}m$, the binary eccentricity suffers very small changes. We also checked whether this assumption is correct on the basis of our simulations. We found that  the  
 mean eccentricity $e$ of the binary IMBH after a 3-body encounter is always 0.70, with a very narrow spread (the standard deviation $\sigma{}$ being always  $1-5\times{}10^{-2}$), and the maximum post-encounter eccentricity is always $\lesssim{}0.8$, in agreement with our statement. 
\\From Figure~6 we note that $t_{gw}$  is longer than $t_{hard},$ 
for $a\gtrsim{}0.5-0.3$ AU.
This implies that the lifetime of a binary IMBH is controlled by 3-body
scattering on 
the hardening time $t_{hard},$ as long as the binary separation  
$a\sim{}$0.1 AU. Therefore, we can neglect the effect of
gravitational wave emission, since we consider binaries wider
than 0.1 AU.

The expected number of supra-thermal stars is a fraction $f$ of the total 
number $N$ of
stars strongly interacting with a binary IMBH. The latter is given by 
(cfr. CMP):
\begin{equation}\label{eq:eq3}
N=\frac{(M_1+M_2)}{m\, \xi{}_E}\ln{\left(\frac{a_0}{a_{\rm {st}}}\right)}
\end{equation}
where $a_0$ is the initial semi-major axis of the binary and $a_{\rm
{st}}$ is the minimum semi-major axis for the encounter to give a
supra-thermal star.  
In eq. (\ref{eq:eq3}) we
impose $a_0=$ 2000 AU, corresponding to 
the orbital separation below which  the binary becomes reasonably hard to generate
supra-thermal stars, 
and $a_{\rm {st}}=$ 0.1 AU, the orbital separation below which the cross section for 3-body encounters becomes negligible. 
We then calculate the number
of stars that remain bound to the cluster and the number of
supra-thermal stars using the statistics derived in our simulations (see the
percentages given in Table 3,
respectively in first column -for the bound stars- and in fourth
column -for the supra-thermal).  The results are shown in Table 7. The
total number of bound stars is always greater than 400 and the number
of stars stirred up to supra-thermal velocities is always greater than
100, even for the lightest system ($M_1$= 50 $M_\odot{}$, $M_2$= 10
$M_\odot{}$).

\begin{figure}
\center{{
\epsfig{figure=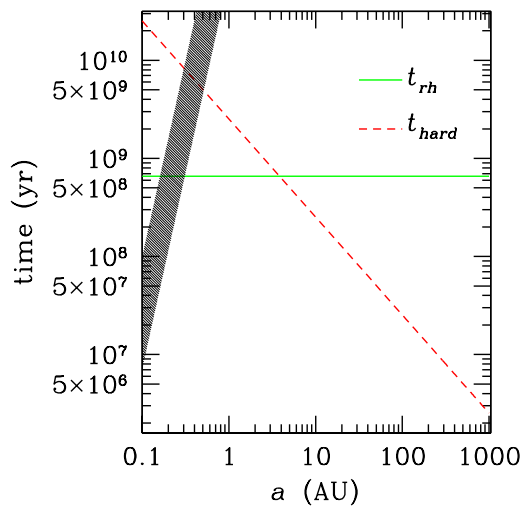,height=8cm}
}}
\caption{\label{fig:fig6} 
  Half mass relaxation time, $t_{rh}$, and hardening
  time, $t_{hard}$, as a function of the initial orbital
  separation $a$ (in AU) of the binary IMBH, for $n=10^5\pc3$,
  $\xi_E=1,$ and $\sigma=8.5\kms$. The shadowed area indicates the
  gravitational wave timescale $t_{gw}$ for binary IMBHs in the
  considered range of masses (from 60 to 210 $M_\odot{}$) and with
  eccentricity $e=0.7$.  }
\end{figure}



\begin{table*}
\begin{center}
\caption{Number of supra-thermal stars$^{\rm a}$
}
\begin{tabular}{llllll}
\hline
\hline
  $M_1^{\rm b}$
& $M_2^{\rm b}$
& Interacting stars$^{\rm c}$
& Bound stars$^{\rm d}$
& Supra-thermal stars
& Supra-thermal stars (last 2 Gyrs)$^{\rm e}$\\
\hline
50  & 10 &  728 &  467 &  128  & 86\\
50  & 50 & 1651 & 1009 & 219 & 146\\
100 & 10 & 2060 & 1567 & 238 & 157\\
100 & 50 & 1722 &  727 & 262 & 158\\
200 & 10 & 1945 & 1057 & 306 & 176\\
\noalign{\vspace{0.1cm}}
\hline
\end{tabular}
\begin{flushleft}
{\footnotesize
$^{\rm a}$ These numbers are calculated for binaries having $a_0$= 2000 AU and $a_{st}$= 0.1 AU\\
$^{\rm b}$ In units of solar masses. $M_1$ and $M_2$ represent respectively the mass of the more and of the less massive black hole in the binary IMBH.\\
$^{\rm c}$ Total number of stars which have an interaction with the binary IMBH.\\
$^{\rm d}$ Number of stars which after the interaction remain bound to the cluster.\\
$^{\rm e}$ Supra-thermal stars formed in the last 2 Gyrs.
}
\end{flushleft}
\end{center}
\end{table*}


Supra-thermal stars may be recognized 
from their proper motion and/or Doppler line shift.  
 Hence we have to take into account projection effects of velocity
 vectors. These effects depend in turn on the orientation of the angular momentum of the binary IMBH $\vecJBH$ with respect to the line of sight.  We find that about 80\% of the supra-thermal stars can be 
recognized as such in the best case, i.e.
when the line-of sight happens to be nearly parallel to
the orbital angular momentum of the binary IMBH $\vecJBH$ (if we are
measuring proper motions) or when the line-of-sight is nearly
perpendicular to $\vecJBH$ (if we are measuring Doppler-shifts). 
On the contrary, the percentage reduces 
to about 50\% when we 
are in the most unlucky cases, i.e. when the
line of sight happens to be nearly perpendicular to the orbital
angular momentum of the binary IMBH $\vecJBH$ (if measuring
proper motions), or when the line of sight
is parallel to $\vecJBH$ (if measuring Doppler shifts). 
Assuming these percentages, we find that the number of recognizable
 supra-thermal stars over the entire cluster is $\sim{}$60-100 for a
 ''light'' binary IMBH ($M_1$= 50 $M_\odot{}$, $M_2$= 10 $M_\odot{}$)
 and $\sim{}$130-210 for a ''moderately massive'' binary IMBH ($M_1$=
 100 $M_\odot{}$, $M_2$= 50 $M_\odot{}$).


 Our calculation refers to  
all stars, including also compact
remnants (neutron stars, white dwarfs) and
stellar types which are too faint for allowing a measurement of proper motion or radial velocity.
Using the numerical code which will be
described in the next Section, we find that the stars with mass from
0.6 to 0.9 $M_\odot{}$ (a mass range which includes red giant (RGB), 
horizontal branch (HB) and bright enough main sequence (MS) stars) represent
about the 45\% of the stars
enclosed within\footnote{We consider the region within 0.1
$r_c$, because we are interested in those stars which have the highest
probability of interacting with the central binary IMBH.} 0.1 $r_c$ of a GC like
NGC~6752. This means that, in the case of a binary of (100$\msun$ , 50$\msun$),
we are able to recognize only 50-100 supra-thermal stars.

 There is a further problem concerning the  number of surviving  
supra-thermal stars. 
Nearly all the proposed mechanisms of formation of binary IMBHs in
 GCs (Miller \& Hamilton 2002; Sigurdsson \& Hernquist
 1993) predict that such binaries are born within the first Gyr since
 the formation of the GC itself. If this is true, considering the
 hardening time $t_{hard}$ plotted in Figure 6, the binary IMBHs which
 still survive today must have orbital separation $\lesssim{}1$ 
 AU (for a  cluster with  $n=10^5$ pc$^{-3}$). 
This means that all the supra-thermal stars produced when the
 binary had orbital separation $a\gtrsim{}3$ AU have already been
 thermalized, because a time much longer than $t_{rh}$ has elapsed
 (Fig. 6). Then, we can observe only the supra-thermal stars produced
 in the last $\sim{}2$ Gyr (about 3 $t_{rh}$, because 
a supra-thermal star can have apo-center distance larger
 than the half mass radius and, therefore, relaxation time longer than
 $t_{rh}$).  Luckily, the orbital separation of a IMBH binary remains
 in the range 0.1-3 AU for a long time, because the hardening rate 
 is lower for such a hard binary, and a large fraction of supra-thermal
 stars ($\sim{}$60\%) is produced in this stage. The number of these
 ''surviving'' supra-thermal stars is reported in Table 7, 6th
 column. Let us consider again the IMBH binary of (100,50)
 $M_\odot{}$. The total number of supra-thermal stars produced by this
 binary in the last 2 Gyrs is $\sim{}$160. This means that, taking
 into account projection effects and considering only bright enough stars 
(i.e. in the mass range 0.6-0.9 $M_\odot{}$), we are finally left with only
 30-60 recognizable supra-thermal stars.

 In this discussion we have not considered the effective instrumental errors so far. We now briefly report on them, without entering in details.
Even if STIS is no more operative, its accuracy remains a good lower limit for future spectrographs. Observing with STIS stars in a globular cluster with distance from the Sun of the order of 4-10 kpc , one can expect an error of 1-2 km s$^{-1}$ in the determination of radial velocity. For example, van der Marel et al (2002), reported spectra of about 130 stars in the core of M15 (distance from the Sun about 10 kpc) with an observational error of the order of 1.3 km s$^{-1}$. 
A somewhat higher error can be estimated for proper motion measurements with the HST/WFPC2. Drukier et al. (2003) combined two sets of observations with the WFPC2 (one taken in 1994, the second in 1999) for a sample of 1281 stars in NGC~6752, estimating a median error of 0.31 mas yr$^{-1}$ with a mode of 0.17 mas yr$^{-1}$. For the distance of NGC~6752 
this means a median error of $\sim{}$6 km s$^{-1}$ with a mode of $\sim{}$3 km s$^{-1}$, which is still an acceptable accuracy to distinguish supra-thermal stars. Similar considerations hold for HST/ACS (see e.g. Anderson 2002; Anderson \& King 2003).  Since we defined supra-thermal the stars with projected velocity higher than $\sim{}$ 12 km s$^{-1}$ (16 km s$^{-1}$ for proper motion measurements), an error of 1-2 km s$^{-1}$ (3-6 km s$^{-1}$  for proper motion measurements) is sufficient to distinguish  supra-thermal from other cluster stars, in clusters like NGC~6752.
In summary, the main problem in detecting supra-thermal stars is not the error on the single measurement but the possibility of observing a sufficient large sample of stars, since supra-thermal stars are expected to be a very small fraction of cluster stars.


\subsection {Spatial distribution of the supra-thermal stars}
\begin{figure}
\center{{
\epsfig{figure=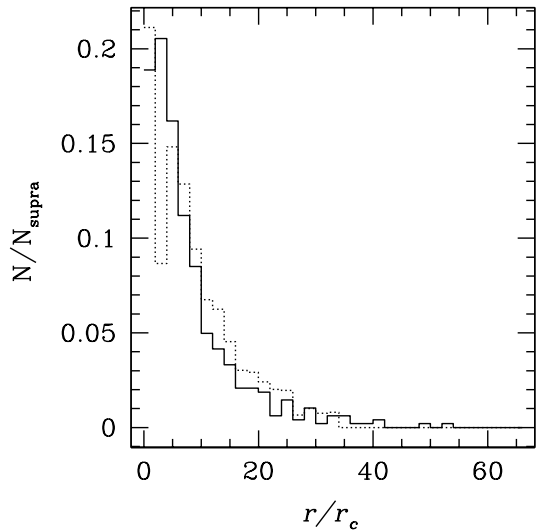,height=8cm}
}}
\caption{\label{fig:fig7} 
Final projected distribution of the high velocity supra-thermal stars ({\it solid line}) in the case D3. The peak is at 3 $r_c$. On the y-axis the number of cases for each bin is normalized to the total number of supra-thermal stars. The ({\it dotted line}) shows the radial distribution of red giant stars in NGC~6752 (normalized to the total number of red giant stars in the sample).}
\end{figure}


 Since supra-thermal stars are only few tens, the possibility of 
recognizing them may be significantly enhanced if their radial 
distribution shows some characteristic
feature. Thus, it is of interest to study how supra-thermal stars
evolve in the cluster and what is their radial distribution.

To this purpose we have explored the dynamics of supra-thermal stars
in a cluster under the action of dynamical friction and the influence
of two-body relaxation effects using an updated version of the code
described in Sigurdsson \&{} Phinney 1995. The code first generates a
cluster background model, i.e. a multi-mass King density profile which
in our case reproduces that of NGC~6752, and we inject in this
background supra-thermal stars whose initial positions and velocities are those
obtained by our 3-body simulations. We followed the dynamical
evolution of these stars for a random time $t$ uniformly distributed
in the range $0<t\leq{}3\,{}t_{rh}$.  In this way we can see how
supra-thermal distribute in the cluster before thermalization. After
evolution for a time t$\lesssim{}3\,{}t_{rh}$ we select the stars that
still are supra-thermal. We applied this procedure to the cases
D1, D2, D3 and D4.

The stars which were still supra-thermal when the simulation stopped
and that can be observed as supra-thermal, taking into account 
two-dimensional (one-dimensional) projection effects and ejections, are
about 70\% (60\%) of the initial sample.  These stars present a radial
distribution like that shown in Figure 7 (for the case D1). This
distribution is peaked around a radius $r_{peak}$ which is of
the order of few core radii for all the considered systems. 
 For a binary IMBH with
$M_1=100\,{}M_\odot{}$, $M_2=50\,{}M_\odot{}$ and $a=1$ AU (D1)
 $r_{peak}=3\,{}r_c$ and we have calculated that it is enough  to measure 
radial velocities within a distance of 6 $r_c$ from the cluster center,
 in order to avoid poissonian fluctuations to smear out the peak.

Taking again into account both projection effects
and luminosity criteria,
 only $\sim{}$30 supra-thermal stars are located within 6 $r_c$.
We found that, if the velocity distribution is
Maxwellian, $\sim{}$10 stars with velocity in the supra-thermal range
are predicted to inhabit within a distance of 6 $r_c$ from the center
of NGC~6752. Then, the number of supra-thermal stars produced by
interactions with a binary IMBH prevails on the
high velocity tail of the Maxwellian distribution. On the other hand
the presence of this tail further dilutes the signature of supra-thermal stars.

In Fig. 7 we have finally compared the radial distribution of 1984 red giant stars (dotted line) observed in NGC~6752 (591 of them from HST/WFPC2 observations, 1393 from the ESO/MPG WFI. See Sabbi et al. 2004) with that calculated for supra-thermal stars (solid line). RGB stars appear to be more concentrated toward the center of the cluster (the maximum being between 0 and 2 $r_c$); but the two distributions are quite similar.

\subsection {Detectability of angular momentum alignment}
We now discuss the observability of the signature
 of the angular momentum alignment effect. 
Column 2 of Table 4 tells us that in the most favorite case about 70\% of the
stars which remain bound are corotating with the binary IMBH, whereas
30\% are counterrotating.  
This means that, if there is a binary IMBH of
$M_1$= 100 $M_\odot{}$ and $M_2$= 50 $M_\odot{}$, we should have about
90-150 corotating and only 40-60 counterrotating supra-thermal stars. 
On the other hand,  angular momentum 
alignment effects are visible  only for
 sufficiently wide binaries ($a>10$ AU). Thus, unless a binary IMBH
is formed recently in a cluster, 
the alignment effect today is completely washed out
 by dynamical friction.

 However, we notice that a very low probability mechanism of
 binary IMBH formation in GCs in recent epochs exists. In fact, there
 is some possibility that a ''last single BH'' (Sigurdsson \&
 Hernquist 1993), ejected from the core in the early stages of the GC
 life, remains in the halo for several Gyrs. The relaxation timescale
 out in the halo is long, and if the orbits can circularize there
 (maybe due to time varying galactic tidal field), then the return
 time to the core is long. When this BH comes back to the core, there
 is a high probability that it forms a binary with the central
 IMBH. Such a binary IMBH would be originated in late epochs. However,
 the probability for this process to occur is low, because it requires
 the BH to receive just the fine-tuned post-encounter velocity needed
 to remain in the outer halo: if the velocity is slightly too high, the
 BH will be ejected from the entire cluster; whereas, if the velocity
 is low, the return time to the core will be too short.

The process described here is an unmistakable feature of a binary IMBH, which 
can not be produced by a distribution of stellar binaries interacting via 
 3-body encounters with cluster stars.
Instead, we need dedicated 3-body simulations to check whether even stellar 
binaries can produce effects like those described in $\S 3.1$ and 
$\S 4.2$. 

\section{Summary}
In our work we explored the effects that the presence of a binary IMBH
imprints on the velocity and angular momentum of cluster stars,
due to 3-body encounters with it. The main results of our analysis can
be summarized as follows.
\begin{enumerate}
\item{} A binary IMBH generates a family of supra-thermal stars,
i.e. of stars which remain bound to the cluster, but have a
velocity higher than the dispersion velocity.
($\S$ 3.1).
\item{} A fraction of stars tend to align their angular momentum,
$\vecJstar{}$, with that of the binary IMBH, $\vecJBH$. This fraction
depends on the reduced mass $\mu{}$ and on the semi-major axis $a$ of the binary IMBH,
but it is always of the order of 55-70\%. This means that the angular
momentum distribution of stars that have suffered 3-body
interaction with a binary IMBH presents a slight anisotropy ($\S$ 3.2).
\item{} The present theoretical models of binary IMBH formation
indicate that these systems are produced in the first Gyr of the GC
life. This means that today, in dense clusters like NGC~6752, binary
IMBHs have, due to hardening, orbital separation $a\lesssim{}$ 1
AU. As a consequence, supra-thermal stars produced when the binary was
wider (more than 2 Gyrs ago for the case of NGC~6752) have now already 
thermalized. 
That reduces the current population of residual supra-thermal stars about of
a factor two with respect to the total population produced during the entire
 IMBH lifetime ($\S$ 4.1). On the other hand these ''surviving'' supra-thermal 
stars
do not show any signature of angular momentum alignment ($\S$ 4.3), because 
this effect is evident only when the stars interact with a relatively wide
binary ($a>10$ AU).
\item{} How many supra-thermal stars are expected to be produced by a binary
IMBH and still surviving in a GC like NGC~6752? 
We estimated few hundreds of supra-thermal stars. Once subtracted the 
fraction of compact remnants and faint stars and taking into account 
projection effects and thermalization of the oldest supra-thermal stars 
(see \S 4.1) we calculate that there may be at most few tens of supra-thermal 
stars suitable to be recognized as such in the whole cluster.
It can be interesting to notice that there is already a claim for the 
detection of high velocity stars in 47~Tuc (Meylan, Dubath \& Mayor 1991).
 \item{} 
Given the aforementioned relatively small number of expected supra-thermal
stars it is important to study their spatial distribution in order to improve
 the chances of recognizing their origin. 
To this
purpose (see $\S$ 4.2), we followed the dynamical evolution of supra-thermal stars in
the cluster potential (assuming a cluster model which reproduces the
observational characteristics of NGC~6752), taking into account
dynamical friction and two body relaxation. We found that, before
thermalization, supra-thermal stars tend to cluster within 6 $r_c$. 
\item Although we have not performed a detailed simulation of the 
observability of the few tens of expected supra-thermal stars with present 
detectors, simple considerations (see \S 4.1 and \S4.2) suggest that the 
various observational biases (mainly related with the limited field-of-view 
of the most sensitive instruments) may lead to the detection of only 
a  sub-sample of the population.
\end{enumerate}

In the light of these findings, the search of supra-thermal
stars and of anisotropies in their angular momentum distribution can
hardly provide useful indications on the existence of binary
IMBHs in GCs with present instruments. However, 
it may become a feasible task for future instruments having much bigger 
collecting 
area and equipped with detectors having much larger field of view, like
(Gilmozzi 2004) the Thirty Meter Telescope (TMT) or the OverWhelmingly
Large Telescope (OWL).

\section {Acknowledgments}
We thank Francesco Ferraro for useful discussions on the issue of
supra-thermal star detection. We also thank Marc Freitag, Cole Miller 
and the anonimous Referee for a critical reading of the manuscript.
 This work found first light during the workshop ''Making Waves with
Intermediate-Mass Black Holes'' (20-22 May 2004, Penn State
University). 
M.~C. and M.~M. acknowledge the Center for Gravitational Wave
Physics for kind hospitality. S.~S. acknowledges support by the Center for
Gravitational Wave Physics funded by the NSF under cooperative
agreement PHY 01-14375 and NSF grant PHY 02-03046. M.~C. and A.~P.
acknowledge financial support from the MURST, under PRIN03.

\appendix
\section{Choice of impact parameters}
We chose impact parameters in order to select all the interactions which have not negligible energetic exchange (about $\Delta{}E_{{\rm BH}}/E^{in}_{{\rm BH}}\gtrsim{} 10^{-3}$), to allow a complete coverage of all the interactions that enter in the supra-thermal domain (20-40 km s$^{-1}$).
We used the formula of the gravitational focusing (Sigurdsson \& Phinney 1993), for which
\begin{equation}\label{eq:eqapp}
b_{max}\sim{}\left(\frac{2\,{}G\,{}M_T\,{}p}{v_{\infty{}}^2}\right)^{1/2},       \end{equation}
where $b_{max}$ is the maximum impact parameter, $M_T$ the total mass of the binary, $v_{\infty{}}$ the relative velocity at infinity and $p$ the distance of closest approach. 

We firstly imposed $p\sim{}(10\,{}a)$ for all the systems (where $a$ is the semi-major axis of the binary). Then we made some test-run to check what were the best maximum impact parameters to include all the strongly interacting stars (and then the supra-thermal stars).
The results of these checks induced us to slightly modify $b_{max}$ as calculated with eq. (\ref{eq:eqapp}). 
In fact, when a binary has a very large semi-major axis (case with $a$=100 or 1000 AU), the choice $p\sim{}(10\,{}a)$ means that the incoming star will pass very far from the center of mass of the binary, will interact very weakly with the binary (unless it closely approaches the secondary component of the binary itself) and never obtain $\Delta{}E_{{\rm BH}}/E^{in}_{{\rm BH}}>10^{-3}$. The velocity at infinity peak is always $\ll{}$ 10 km/s. 
 
On the other hand,  when the binary has a very small semi-major axis ($a\leq{}$1AU), 
the choice $p\sim{}(10\,{}a)$ does not take into account all the strong interactions. In fact all the stars which have $p\sim{}(10\,{}a)$ pass so closely to the very hard binary to become very fast ($>$ 100 km/s). But, if we consider only $p\sim{}(10\,{}a)$, we miss most of the interacting stars. In fact, if we take $p\sim{}(50-100)\,{}a$, we still have stars that sufficiently approach to the center of mass of the binary to receive an outgoing velocity $\sim{}$ 20-40 km/s (exactly the range of supra-thermal stars). And, statistically, stars with $p>10\,{}a$ will be much more numerous than stars with $p<10\,{}a$. Then, if we take  $p\lesssim{}(10\,{}a)$, we miss the bulk of interacting stars belonging to the supra-thermal interval. 

Then, both when we choose a small $b_{max}$ ($p\lesssim{}10\,{}a$) for a very hard binary and a large $b_{max}$ ($p\gtrsim{}10\,{}a$) for a wide binary, we introduce a bias. In the former case we select only high velocity encounters ($\Delta{}E_{{\rm BH}}/E^{in}_{{\rm BH}}\gg{}10^{-3}$), omitting the bulk of interacting stars; in the latter we risk to consider also unperturbing flybies ($\Delta{}E_{{\rm BH}}/E^{in}_{{\rm BH}}\ll{}10^{-3}$).

\begin{figure}
\center{{
\epsfig{figure=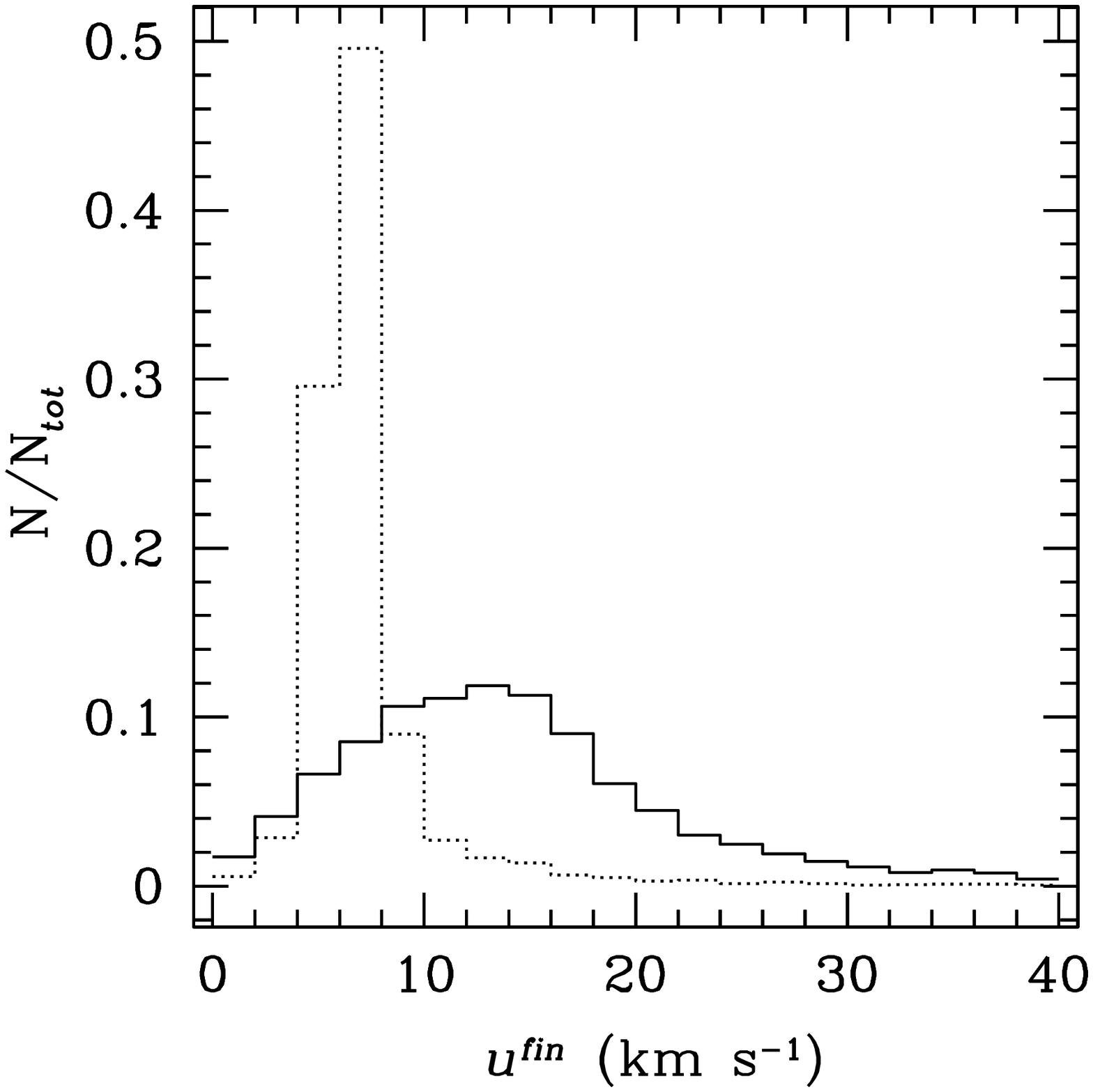,height=8cm}
}}
\caption{\label{fig:fig8} 
Post-encounter asymptotic velocity distribution of the star, after the interaction with a binary of $M_1$=50 $M_{\odot{}}$, $M_2$=10 $M_\odot{}$, $a$=100 AU. The {\it solid line} shows the case with $b_{max}$=100 AU (the runs reported in the paper as case A3), whereas the {\it dotted line} shows a case with $b_{max}$=2000AU (corresponding to adopt $p=10\,{}a$ in  eq. (\ref{eq:eqapp})).  On the y-axis the number of cases for each bin is normalized to the total number of
resolved runs (N$_{tot}$).}
\end{figure}

\begin{figure}
\center{{
\epsfig{figure=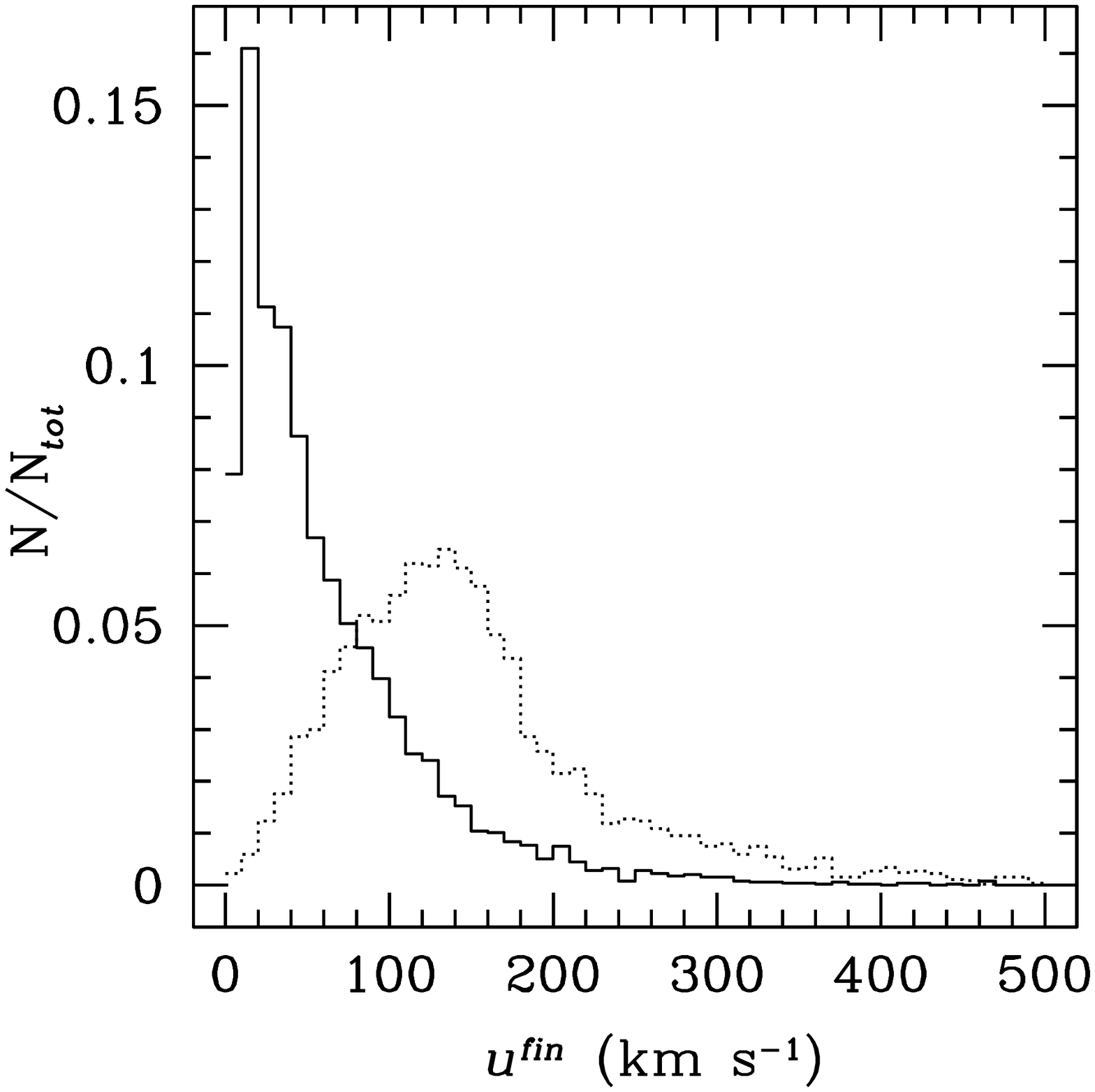,height=8cm}
}}
\caption{\label{fig:fig9} 
 Post-encounter asymptotic velocity distribution of the star after the interaction with a binary of $M_1$=50 $M_\odot{}$, $M_2$=10$M_{\odot{}}$, $a$=1 AU. The solid line shows the case with $b_{max}$=60 AU (the runs reported in the paper as case A1), whereas the dotted line shows a case with $b_{max}$=1 AU (corresponding to impose $p=1\,{}a$ in  eq. (\ref{eq:eqapp})).  On the y-axis the
number of cases for each bin is normalized to the total number of
resolved runs (N$_{tot}$).}
\end{figure}


Fig. A1 represents the post-encounter asymptotic velocity distribution of the star in the case of a binary with $M_1$=50 $M_{\odot{}}$, $M_2$=10 $M_\odot{}$, $a$=100 AU. 
The dotted line shows the case with $b_{max}$=2000 AU (corresponding to adopt $p=10\,{}a$ in  eq. (\ref{eq:eqapp})):  most of interactions are very weak flybies (velocity $<$ 8 km/s), and we cannot count them as strong interactions.
Instead, the solid line shows the case with $b_{max}$=100 AU ($p=a$), i.e. the runs reported in the paper as case A3: this represents a valid statistic sample of all the strong interactions (the peak being at $\sim{}$ 13 km/s).\\
On the other hand, Fig. A2 shows the post-encounter asymptotic velocity distribution of the star in the case of a binary with $M_1$=50 $M_\odot{}$, $M_2$=10$M_{\odot{}}$, $a$=1 AU. 
 The dotted line, showing the case with $b_{max}$=1 AU (corresponding $p=1\,{}a$), is peaked at very high velocities. 
Instead, the solid line, for $b_{max}$=60 AU (case A1), shows that, if we take into account also encounters with $p>1\,{}a$, but still with $\Delta{}E_{{\rm BH}}/E^{in}_{{\rm BH}}>10^{-3}$, we have a very different velocity distribution, peaked at 20-30 km/s. We think that the last one is the only statistically significant case, because it takes into account all the interacting stars with significant energetic exchange. The dotted line, instead, misses the most numerous class of interactions.

For these reasons we adopted new criteria for the choice of $p$ (to be substituted in eq. \ref{eq:eqapp} to derive $b_{max}$) as:
\begin{itemize}
\item[-]if $a\geq{}$100 AU, then $p\sim{}a$;
\item[-]if 1 AU$<\,{}a\,{}<$100AU, then $p\sim{}(10\,{}a)$;
\item[-]if $a$=1 AU, then $p\sim{}(60-100)\,{} a$.
\end{itemize}
This means, that our selected maximum impact parameter is quite constant ($\sim{}$100 AU), independent of the semi-major axis a of the binary (at least if $a<$ 1000 AU).



We think that the big 
maximum impact parameter required by systems with $a$=1 AU
is due to the uncommonly large mass ratio between the binary and the
incoming star. In fact, when at least 
one of the two components of the binary is so massive, it plays an 
important role on the gravitational focusing, nearly independently of the 
semi-major axis. Then, the incoming star is attracted toward the center of 
mass of the binary, even if it starts with very large impact parameter. 
But a binary of $a$=1 AU has a very small geometrical cross-section; then 
only a very small fraction of interacting stars will have an effective 
pericenter $p\leq{}a$. Thus, only a few stars will interact very strongly with 
the binary. Most of the stars will have a pericenter $p>a$; but 
the binary has a so large binding energy that also these stars receive 
non-negligible post-encounter velocity, and they can become supra-thermal.

Even the behavior of $\xi{}_E$ shown in Fig. 2, apparently different from what predicted by the Heggie's law when $a=$1 AU, depends on our choice of $b_{max}$.  
 In fact, if for a binary with semi-major axis $a=$ 1 AU we choose $p\sim{}a$ (corresponding to $b_{max}\sim{}1$ AU, instead of 60-100 AU), we obtain $\xi_E{}\sim{}4$, higher than the value of $\xi_E{}$ derived for wider binaries and in agreement with Heggie's law. On the other hand, we need to select $p\gg{}a$, because we want to include all the interactions which have not negligible energetic exchange. But this choice lowers the average value of $\xi_E{}$ from 4 to about 1.



\newpage
\newdimen\minuswidth    
\setbox0=\hbox{$-$}

\newpage

\end{document}